\documentclass[12pt]{article}
\usepackage{amsmath}
\usepackage{graphicx,psfrag,epsf}
\usepackage{enumerate}
\usepackage{natbib}
\usepackage{url,breakurl}
\newcommand{\blind}{0}

\date{14 Apr 2022}

\begin{document}\sloppy

\def\spacingset#1{\renewcommand{\baselinestretch}%
{#1}\small\normalsize} \spacingset{1}


\if0\blind
{
  \title{\bf Delivering data differently\thanks{The authors gratefully acknowledge Jessica Hullman, Alberto Cairo, Paul Murrell, Kim Kowal Arcand, Mitzi Morris, Thomas Lumley, Martin Wattenberg, and Dmitri Tymoczko for helpful comments.}}
  \author{gwynn sturdevant\hspace{.2cm}\\
Laboratory of Innovation Science, Harvard University
    \and
A. Jonathan R. Godfrey \\
School of Mathematical and Computational Sciences, Massey University
    \and
Andrew Gelman \\
Departments of Statistics and Political Science, Columbia University}
  \maketitle
} \fi

\if1\blind
{
  \begin{center}
    {\LARGE\bf Title}
\end{center}
  \medskip
} \fi

\begin{abstract}
Human-computer interaction relies on mouse/touchpad, keyboard, and screen, but tools have recently been developed that engage sound, smell, touch, muscular resistance, voice dialogue, balance, and multiple senses at once. How might these improvements impact upon the practice of statistics and data science? People with low vision may be better able to grasp and explore data. More generally, methods developed to enable this have the potential to allow sighted people to use more senses and become better analysts. We would like to adapt some of the wide range of available computer and sensory input/output technologies to transform data science workflows. Here is a vision of what this synthesis might accomplish. 

\vspace{\baselineskip}

\end{abstract}

\vspace{\baselineskip}
\spacingset{1.2}

\noindent%
{\em ``Part of the aesthetic orientation is a perceptual openness, a willingness to notice and believe in connections and meanings that may not be instantly apparent.''} --- Elizabeth Hellmuth Margulis, {\em On Repeat}.

\section{Introduction}
\label{sec:intro}

In traditional Andean South America, people recorded data using knotted strings called khipu. After their arrival, the Spanish incorporated khipu to both ecclesiastical and administrative purposes \citep{BrokawGalen2010Ahot}. Like many tools that were once useful, knots are no longer used to store or communicate data in South America.

Another data visualization tool was invented fifty years ago to display multivariate data using a coding that produced cartoon faces \citep{Chernoff}. Because a large portion of the human brain is reserved for processing facial features, scholars believed that this presentation made it easier for humans to understand data. The practical value of this idea did not match its theoretical appeal.

These ways of delivering data are of interest because they are not widely used. What are the similarities between these data delivery methods and other methods that are not used? What data delivery methods have recently been advanced that will not be used in twenty years? And most importantly, what methods will be used?

From the other direction, some familiar visualization tools are surprisingly recent. For example, the scatterplot appears to be less than two hundred years old, appearing centuries after the development of Cartesian coordinates; see  \cite{HistScat}. We can consider this relatively late appearance as frustrating or as a source of hope that perhaps other valuable tools remain waiting to be developed.

\subsection{From visualization to vivification}

In this article, the authors consider a number of ways that data can be delivered differently. These methods have the potential to enhance data delivery for all people, and could be monumental for people with low vision for whom literal visualization of data is difficult or impossible. 

Terminology is important, and the authors have spent time searching for words that encapsulate this work. Unfortunately, we have not found any single ideal term. Candidates have been visualize, visceralize, sensify, Umweltbeziehung, Umwelterfahrung, and vivify. Visualize does not include the multiple senses that we hope to engage and may not be relatable to people with low vision. Visceralize  has strong resonances of ``viscera,'' the internal organs of the body, not necessarily just the senses. The root word of sensify is sense and we will not be limited to the five senses that are traditionally considered. German has words for everything. Exploring resulted in Umweltbeziehung or Umwelterfahrung which are challenging for English speakers. Although nothing actually comes to life, for this article the authors have settled on vivification.

Related is the biosemiotic term Umwelt, first used by Jakob von Uexküll \citep{Sebeok1976,umwelt}. The Encyclopaedia Britannica defines Umwelt as ``perceptual environment'' and the Oxford English Dictionary defines it as ``The outer world, or reality, as it affects the organisms inhabiting it.'' Each organism experiences the world through a perceptual bubble. Dogs experience much of the world through scents. Humans generally use sight as our primary cue. When you go for a walk with your dog, your Umwelts are completely different due to your differing senses. The movie Superpower Dogs uses visual cues to present the dog Umwelt to humans.

Umwelts differ within as well as between species. When a biologist and a geologist go for a walk, they can have completely different Umwelts due to different education and interests. One impact of different Umwelts is differing internal experiences; two people may have different imaginations of a box. To accommodate and supersede these differences scientists could consider delivering data differently.

The authors can formally define data vivification to mean a mapping of a dataset to the human Umwelt. Historically, statisticians have focused largely on linking visual representations of data to the brain. We propose that a multitude of other representations can assist the data pipeline. These include auditory, tactile, and muscular resistance senses.  As \cite{wilkinson_vivification} put it,  ``The graphs we have been talking about cannot be seen, heard, tasted, or otherwise perceived. It is the job of the Aesthetic object to act as a function on the elements of a CGraph and translate them into real numbers or characters that control some display device, such as a CRT, plotter, sound generator, or even odor generator. Once we have done this, we have what I call a graphic. A graphic is a perceivable graph.''

The options presented in this article are restricted to today's thinking and must surely be a subset of what a wider collective might dream up. To that end, the authors think of this work as a conversation starter, and it is a conversation worth having, and worth involving as many people as possible, among engineers, developers of statistical methods, and users of statistics. At times, the only difference between an awesome novelty and a foolish notion is the uptake by users; we need inventiveness to create innovation. Until we know how well received an innovation ends up being, all ideas are only potential solutions and opportunities.

\subsection{Errant backward compatibility}

A fable shared on the internet discusses backward compatibility and rocket ships \citep{horse_rockets}. In this story, the first roads were designed for Roman chariots pulled by two war horses. With this standard established, subsequent transportation used the same gauge without reflection. Eventually, this lack of reflection had consequences. Boosters for rockets to be launched in Florida were built in Utah. Transportation from Utah was a limiting factor for the construction of these boosters; the engineers were limited because of the dimensions of the mountain tunnel through which the train carrying the rockets passed. Because of backward compatibility, the width of war horses impacted upon space travel.

This fable illustrates the need for careful reflection on mundane details---or perhaps its popularity demonstrates a desire for a tidy explanation for our organizational oversights.

Human-computer interactions mostly happen via mouse, keyboard, and screen. The mouse mimics infant communication by pointing, while screens and keyboards are largely modeled on paper-based communication. Emails have the same format as letters but travel faster, and spreadsheets have a similar format to Babylonian tablets. Most computer-generated data visualizations are digital updates of paper-based formats. What is a computer's full capacity to support human interaction with data?
 
\subsection{The overlap between truth and beauty}

Statistical vivification falls in the intersection of science, engineering, and art.  The authors have practical and statistical concerns about data display and model understanding; how do these connect to questions of graphical beauty?  An apparently hard-nosed realistic response would be that, in engineering, there is no beauty beyond utility.  But we think that would be too glib, for several reasons.  First, aesthetic concerns have in the past often led, rather than followed, practical developments; consider, for example, all the scientists and software designers who have been influenced by the ideas of \cite{Tufte1983} and \cite{Tufte2006}.  Second, as noted by \cite{hansen2013}, choices of display are also questions of design, and the same principles that can make data-based art appealing \citep{ListeningPost} can inform functional vivifications, in the same way that innovative or ``cool'' engineering can facilitate scientific development.

\section{Tools for conveying information through different senses}
\label{sec:tools}

\subsection{Sight}
Data visualizations are the most common mapping between data and the human Umwelt. Since the 1970s, scholars have developed both static and dynamic interactions of this type. Trellis plots are still common but rotating 3D plots have fallen out of use. The Grammar of Graphics \citep{gg} continues to have a big impact on the current visualization landscape as implemented at Tableau and the R package ggplot2 \citep{ggplot2}. New ideas have been developed to visualize large datasets \citep{Unwin2006}.

Data visualization bridges between statistics, graphic design, and software engineering. Information visualization and statistical graphics have different goals, but both have become an important part of our culture with news outlets displaying them throughout their websites \citep{InfoVis}.

An extension of data visualization is immersive analytics, using virtual reality headsets and hand sensing devices to explore data. Some of these are FIESTA \citep{9222346}, VRIA \citep{8954824}, Uplift \citep{Uplift}, u2vis \citep{PARIV}, and Tiltmap \citep{yang}.

The authors see visualization as a starting point when thinking about other vivifications. In particular, the large body of work on research and practice in visual communication should inform progress in the theory and practice of other sorts of  vivification. 

\subsection{Touch}
3D printing and tactile graphics are a direct way that people can feel data. Some methods involve Braille printers \citep{Braillo,APH}, embossing \citep{Viewplus}, and contouring \citep{NBP}. Many of the available tools are open source \citep{rayshader,TactileR,RGL}. All can be improved. As 3D printings are durable, it makes sense to improve the accessibility of this technology by adding descriptions about axes, tick marks, point characteristics, and other assumed aspects of plots.

\cite{GoP} maintain an online gallery of physical visualizations, including a 2010 adaption by Hans Rosling who uses Ikea boxes, pebbles, fruit juice, and even toilet paper. The first ever interactive data visualization was physical. \cite{bertin_1969} developed a way to cluster data using physical cubes arranged in a matrix where columns and rows can be interchanged. His method was groundbreaking since computational methods to cluster complex data did not then exist.

Because the production cost of that physical representations of data is decreasing while the fidelity and flexibility continues to increase, some scholars believe physical representations ``will eventually support data analysis tasks as complex as performed on today's desktop computers'' \citep{dataPhys}.

\subsection{Muscular resistance sense}
Researchers are in the early stages of developing haptic technologies and the authors see potential to use this technology to vivify data. Haptic interfaces currently include gloves \citep{gloves}, hand-tracking devices \citep{handtrack}, recognition of mid-air shapes \citep{shapes}, and vests \citep{Vest}. Some scholars have developed tools which allow the public to ``exploranate'' the nano scale \citep{8994161}. It is even possible to build your own haptic equipment that detects finger movements \citep{DIY_haptic}. One promising technique is used by physicists to allow participants to visceralize dark matter as cave formations \citep{cavideo}. There is open-source technology that we believe will be useful in extending haptics to statistical tools \citep{Bocelli, protohaptic, FD}.

MIT's Artificial Intelligence Laboratory has developed tools for virtual tactile sense communication. Their interface allows users to alter ``shape, texture, temperature, weight and rigidity \dots through a device called the PHANToM haptic interface'' \citep{Phant}. That particular haptic interface is currently produced by SensAble Technologies; Immersion produces a competing product called TouchSense.

\subsection{Balance and kinesthetic sense}
The Wii technology—motion sensors in the form of a small remote (``Wiimote'') which allow a user's movements in three dimensions to be conveyed to a computer, and which also allow the computer to send feedback to the hands—has already been extended beyond sports video games (golf, baseball, and the like) to assist in rehabilitating stroke victims \citep{vassou2008} and in physical therapy sessions \citep{drummond2008}. At the moment the authors are not entirely sure how the Wii technology would be adapted to statistical analysis and data visualization, but we expect that there are some good possibilities and applications to be discovered or created.

\subsection{Sound}
During the Cold War, \cite{speeth1961} used audification techniques to discriminate between a bomb and an earthquake and had a 90\% detection rate. He expected his methods to be quickly adopted but was unfortunately rejected and isolated, and his security clearance was retracted \citep{AxelVolmar2013LttC}.

Anyone who has heard an unusual rattling on their bike understands the usefulness of audification. Additionally, mechanics' ability to diagnose problems based upon sound demonstrates that refining listening skills can enhance our understanding of audified data. The authors also know of instances where people with low vision have been able to diagnose issues with wheel alignment in a car by touching the steering wheel (they were not driving).

Pioneering software engineer Margaret Hamilton told the following story \citep{Creighton} of the SAGE project from around 1960:
\begin{quotation}
\noindent
``When my program was running, it always sounded like the sound of waves on a beautiful seashore, so we all referred to it as the `seashore program.' Until one night at 4 am in the morning when I got a call from one of the computer operators who said something terribly wrong has happened with your program. When I asked him how he knew, he said ‘it no longer sounds like a seashore!' Now, we had a new way to debug, using sound!''
\end{quotation}
This episode indicates how our brains' ability to distinguish timbre can be used to detect changes in data.

Attempts to use sonification for data have been studied since \cite{PIIS}, and there are many directions for improvement: in addition to being beautiful, information can also sound beautiful. Currently, some astronomers develop beautiful sonifications \citep{sonification} but the laborious process is not yet automated or available in multiple programming languages. Scientists have built censuses of sonifications \citep{SonificationCensus}. Existing open-source software can be used as a starting point \citep{ThinkDSP, sonify, tuneR, Highchart, Twotone}. Using musical skills, the authors plan to improve and simplify this process.

There are multiple facets to auditory display \citep{2011Tsh}, including exploratory data analysis, parameter mapping sonification, model-based sonification, parametric, and nonparametric methods.

In heliophysics, audification helped researchers identify ``artiﬁcial, instrument-induced noise that was not previously observed by the scientist and also the identiﬁcation of wave activity embedded within turbulent solar wind data'' \citep{Audification2014}. Two months later the researchers continued use of this technique.

Positions of points on a scatterplot and lines on a graph could, with care, be indicated with sound.  Compared with the senses of touch or body location, sound is less localized, so it would make sense to use pings not to tell where data are but rather when they are encountered.  For example, if a robot arm (or, perhaps better still, the user's bare hand) is swept over a constrained space, data thus encountered on the virtual scatterplot could make the sound of rain, with individual drops or pings for isolated datapoints and a louder pitter or torrent for increasingly dense clusters.  This method was used to audify COVID-19 deaths \citep{soundC19}. The authors could also incorporate spatialized 3D sound technology in this area.

More generally, sound could be used at the user's discretion to reveal the values of additional variables that are not in the physical display.  For example, the user could click a mouse at any time to learn additional information about any data point that he or she encounters in exploration, or the computer could be asked about information such as the number of nearby points, the integral of a function in a specified range, or other summaries of a graph.

\subsection{Voice dialogue}
Voice driven development is another technological tool that can increase accessibility \citep{NowogrodzkiAnna2018Sich, HBSVDD}. There is software that assists those with repetitive strain injury that can no longer type with ease to ``speak'' with their computer \citep{IEEEVDD, Talon, Serenade}. The software allows users to verbally code in computing languages \citep{VDD_python}. We will build something more conversational where the computer recognizes keywords and fills in a template with variable names. The authors show a hypothetical example script in Section \ref{sec:UF}.

The DataBreeze interface currently ties together pen, touch, and speech-based interactions to explore data \citep{9023002}. This type of technology could be made available in statistical software.

Another way to interact with visualizations is natural language interfaces. Eviza \citep{Setlur2016} is a Tableau prototype that allows users to interactively dialogue with an existing visualization. Users do not need to know Tableau; the phrase “show me wildfires in California” will execute a query. DataTone uses voice recognition software to the address multiple ambiguities when querying a database \citep{NLPAmbi}. For example, the above query does not specify the year or the size of the wildfire. DataTone will display boxes from which the user can select both year and size and updates the plot as the user makes their selections. FlowSense is a visualization tool that also has voice interaction and allows users to type queries into the online tool \citep{Yu}. 

The authors do not expect voice interaction to supplant the usual menu or command-based systems.  Oddly enough, however, many of the steps that will be required for the voice dialogue system---an ability to anticipate the user's requests and to present results concisely---are important in statistical software in general.  It may be that the tools we develop here will prove useful generally as well as for those who cannot type for reasons relating to their dexterity.

\subsection{Smell and taste}

In 1960 Elizabeth Taylor starred in The Scent of Mystery, which involved not only audio and visual stimulus but novel technology called Smell-O-Vision; it was the first ever movie that stimulated olfactory senses beyond the smell of buttered popcorn. The company Scentevents participated in a remake \citep{Scentevents}. Scratch and sniff cards have also been used for a similar purpose for the movie Polyester \citep{Odor}. \cite{NakamotoTakamichi2016EoMO} offers an overview of current research in conveying information through smell and taste; the author admits that this research is in a primitive state. A chef from a three Michelin-starred restaurant was building an olfactory extension to the iPhone \citep{Scentee}. A company in Vermont is also working to include scents into virtual reality headsets \citep{modern_smells}. In addition to visual cues, stimulation of olfactory senses can be considered to explore data.

\subsection{Combining multiple senses}

In addition to technological advances that deliver data differently, there are also creative methods. An example of this is the Boston Coastlines Future Past in which artists organized a walking tour of the projected coastline after sea levels rise from global warming \citep{BostonCoast}. A Sort of Joy \citep{Joy} is another example of delivering data differently. In this short play, artists with work at the museum are organized by first name. Instead of a bar chart to show which first names are most common, actors spoke the name and allowed time to pass that was related to the height of the bar. Additionally, instead of coloring bar charts based upon gender, male voices verbalized names of male artists and female voices names of female artists. Another example involves data physicalization of water pollution created collaboratively with community partners in Chelsea, Massachusetts \citep{PAR}. Some artists are asking the question, ``What does data theatre look like?''\ \citep{Brea}. The authors believe there are scientists who may be interested in these questions.

A more concrete example of combining senses could be the exploration of level curves of a price function (or, for a more homely example, isotherms in a weather map) using the robot arm.  As the user pushes the arm across the curves, the system could say aloud what is happening (``Crossing the twenty-degree line from below \dots crossing the thirty-degree line from below  \dots'').

\section{Perceptual differences}
\label{sec:perception}

Each sense has a unique interaction with the brain. Most people have favorite songs that they repeat, but it is unusual to reread a book or rewatch a movie. Music cognitionists agree that music is a repeating sense \citep{MargulisElizabethHellmuth2013ORHM}. Music also elicits emotion as a matter of course, unlike visual imagery which tends to elicit emotional reaction only when tied to particular content. Another advantage of a sound is that it may not consume all the attention of the brain the same way that movies, books, or visuals do. Background music is a common term that describes this. In visual senses, focus can impact upon our ability to perceive obvious things even if our eyes see them. One example of this is the invisible-gorilla study \citep{ChabrisChristopherF2010Tig:}.

The authors believe that the processing methods of the brain should impact upon how the senses are used.  Compared to sound and sight, a haptic tool making use of the muscular-resistance sense is more conversational. The tool reacts to user input by giving more information which results in a looping effect similar to speaking with someone.

Unlike some of the other senses, smell has a strong discrete aspect and does not always naturally fall on a continuum; hence the authors suspect it could be used in the way that we use different symbols or colors to distinguish points on a graph, or as an alarm (if the sound channel has already been taken for more useful purposes).  

Taste, like sound, has an emotional valence that responds positively to repetition. Researchers have shown that childhood exposure to high salt and high fat diets ``exacerbate[s] a number of illnesses and conditions in adulthood''  \citep{MedicineInstituteof2010StRS, GIDDINGSamuelS2009IAHA, EkelundUlf2007AoWG, BarkerD.J.PDOoA, MennellaJulieA2014Ootp} which suggests that taste is a sense that is calibrated in early childhood.  It is not clear to us how taste would be used in data perception, but to the extent this is done, we would want to make use of this property.

Touch and the sense of body location are inherently spatial but in a different way than the spatial organization of vision.  Another complexity is the relation between sensory perception in time.  For senses other than sound, observation in time can require active focus on short-term memory.  For example, the dynamic scatterplots of \cite{rosling2015} have been extremely popular as a way of visualizing the way that differences between countries change over time, but it becomes difficult to ``read'' these graphs with a goal of understanding details:  there are too many patterns to focus on at once, and this leads to a frustrating pattern of repeated viewing and attempts at memorization.  The goal is not just to see things as they are happening; we also need to store the patterns, and deviations from these patterns, in an accessible and inviting way.

Designing tools that engage non-visual senses in a statistical workflow can be an overwhelming task. The authors suggest that scientists think carefully about the perceptual advantages of the sense they would like to incorporate and work to discern how the advantages can improve their workflow. Braille is an accessible form of writing for blind and low vision people, but American Sign Language is a completely different language dissimilar to English. Both are necessary. Vivifications of plots for the blind are important, but the authors additionally want to develop a whole language of interactions with computers that will improve statistical workflows. Our goal is beyond replacing visualizations.

\section{Some possible applications to statistics}
\label{sec:examples}

The authors would like to use vivification technology to improve statistical practice, both for people with low vision and for analysts more generally who can use non-visual senses in order to increase their understanding of data and models.  In this section we speculate on different ways in which technological developments can be turned into tools for data analysis.

\subsection{Sensory scatterplots and imaging}

Remember the toy that is made up of pins held loosely in a grid of holes?  You push up with your hand on one side, and then a three-dimensional image of your hand in pinheads appears on the other side.  One could use an electromechanical version of this toy to graph surfaces, curves, and scatterplots in a way that could be felt by the user's hand.  Linked up to a computer,  the pinboard could also convey a series of snapshots or changes over time, thus potentially displaying four dimensions in a way that is not possible using standard visualization tools.  Whether or not the pinboard becomes useful as a supplementary tool for sighted users, the authors anticipate that it could be an excellent way to convey complex graphical information for people with low vision. The American Printing House for the Blind confirmed it is currently developing such a device, and a more expensive European model is currently available.

When 3D printing a plot, variables could be mapped to different textures. With X-ray data, NASA maps silicon to a smooth texture, iron is clumpy, and argon and neon are sandpaper with differing grit \citep{Arcand}. If each variable is mapped to a different texture, the authors would recommend 3D printing the path of each coefficient alone, and the axes with tick marks, and a grid alone. The full plot could be built by stacking all the pieces together which is similar in process to building a ggplot.

\subsection{Voice dialogue for real-time data analysis}
\label{sec:UF}

Voice-driven development combined with natural language interfaces has the potential to benefit data scientists and statisticians. Here is a potential script:

\begin{quotation}

\noindent
User: Regress earnings on height.

\noindent
Computer: Earnings equals sixteen thousand plus four hundred times centered height plus error.

\noindent
User: Error?

\noindent
Computer: Mean zero, standard deviation eighteen thousand.

\noindent
User: Centered height?

\noindent
Computer: Height in inches minus sixty-five.

\noindent
User: One more significant digit.

\noindent
Computer: Sixty-five point two.

\noindent
User: Standard errors of the regression.

\noindent
Computer: Intercept: sixteen thousand, standard error two thousand. Coefficient on centered height: four hundred, standard error one hundred fifty.

\noindent
User: Include sex as a predictor.

\noindent
Computer: Earnings equals four thousand plus eleven thousand if male plus two hundred times centered height plus error.

\noindent
User: Interact sex and height.

\noindent
\dots
\end{quotation}

The idea is that interactive dialogue maps to statistical workflow in a way that is not captured by usual coding or menu-based approaches.  Thus, this could be a tool that would augment the toolkit of sighted data analysts as well as providing a replacement for visualizations for those with vision impairments.

The authors suspect that a voice-recognition program where users can regress on chosen variables, which then speaks out coefficients and errors would greatly improve statistical processes. The user may then ask for regression with interaction terms. This process could be used with linear regression, logistic regression, and include multilevel models with different voices that could then speak coefficients and errors.

One of the main commercial applications of a computer voice dialogue is for it to act as a human or even to fool users into thinking it's human (as for example when it's the back-end for an online tool to resolve customer complaints). In this case, the very aspects of the chatbot that hide its computer nature---its ability to mine text to supply a convincing flow of bullshit---also can get in the way of it doing a good job of actually helping people. This suggests that chatbots would be more useful if they more transparently revealed their computer nature and didn't try to emulate people, in the same way that its better for a chessbot to just be itself and not try to fake us into thinking we're playing against a human.

\subsection{Voice recognition for exploring matching in clinical trials}

\cite{STURDEVANT2021100746} present a tool for finding balance in a two-armed clinical trial at randomization. The authors provide a Shiny web application \citep{Shiny} where researchers can practice randomization conditional on a set of matches. To update the matches, researchers must scroll down to the variable, update the weight, then scroll to the top again to view possible randomizations. Scrolling disrupts the exploration process and locating the desired variable is tedious. One solution to this is voice recognition. After turning on voice recognition, researchers could use the keywords to update weight, then the variable name and the weight to interact with the computer. The user could then click on a button at the top of the ``Matching Variables Info'' tab that tells the application to update the matches and produce a plot.

\subsection{Using musical sounds to convey the progress of iterative algorithms}

Instead of using sound as a replacement for visuals, we can augment our interactions with computers with sound. For example, modern data-fitting algorithms tend to be iterative, with an initial transient period as the algorithm gets close to the solution and then a slower convergence to a point (for an optimization algorithm) or a distribution (for a Monte Carlo algorithm).  In either case, the progress of the algorithm could be expressed using musical sound, with a change in pitch during the transient period, settling to a single note or distribution of notes.  More can be done along these lines: savvy users know to run multiple chains of algorithms such as variational inference or Markov chain Monte Carlo using different starting points.  The spaces in parameter space being occupied by different chains can be represented by different keys and timbres, so that the progress of mixing of initially-separate chains can sound something like different musical instruments being tuned.  

We can also use pitch, volume, rhythm, timbre, and the progressions of musical expectation to convey what is happening within the fits and starts of an iterative algorithm.  For example, Hamiltonian Monte Carlo can get stuck when its steps are too small, too large, or misaligned to the local geometry of the space being explored; within an algorithm such as the no-U-turn sampler, these problems can appear as divergences or max treedepths.  Instead of expressing these as discrete warnings on the screen, these could be conveyed, for example, through pitch, with well-behaving trajectories sounding like calm music, gradually turning into annoying high-pitched buzzing when the algorithm is getting stuck.  This sonic summary could be tied to a dynamic visualization so that the user could then take a look at where in the parameter space this is happening.

A related application is the training of neural networks, where progress is tracked using improvement of some objective function during the training process, and, again, sound can be used to convey progress of the algorithm from different starting points, with unusual or unexpected sound patterns indicating a lack of smooth progress.  The sonic of musical output can serve to reassure that the algorithm is progressing well or alert users of problems, and also to signal when and where to perform further exploration, perhaps using visualizations, to diagnose and fix problems.

\subsection{Exploring fitted curves and likelihood surfaces haptically}

Imagine a robot arm that can be moved within a $30 \times 40$ cm box.  Such an arm could be used by a blind user to ``feel'' a curve (for example, a nonlinear regression or spline, or a mathematical function such the logarithm or the normal distribution curve), as follows: the arm would start at one end of the curve and the user could grip it and move it along, with the motion physically constrained so that the arm would trace the curve.

In displaying several surfaces---for example, level curves indicating indifference curves---the arm could start on one of the curves and be programmed to stay on that curve, unless the user pushes hard, in which case there would be resistance during which the arm moves between curves.  It would then lock into the next curve, which the user could again trace until he or she pushes hard enough to get the arm unstuck again.

More generally, the robot arm could be used for exploring three-dimensional functions such as physical potentials, likelihood functions, and probability densities.  From any point in the two-dimensional box, a ``gravitational force'' would pull the arm toward a local minimum (or, for a likelihood or probability density, the maximum) of the function. Then with moderate effort the user could move the arm around and, by feeling the resistance, get a sense of gradients, minima, and constraints.

Various displays could be implemented as three-dimensional functions and then explored using the robot arm.  For example, a scatterplot can be converted to a kernel density estimate.  When moving across sparse data, the arm would feel a slight pull or stickiness as it moved past individual data points; in data-dense parts of the scatterplot, the massed points would pull the density down while at the same time allowing the user to feel individual data as small local tugs on the arm. For this purpose, it might make sense to use a kernel function such as a Student-t so that each point has a sharp local peak within a broader area where it contributes to the density estimate.

Alternatively, walking with each hand on a physical curve to feel the differences between them is another possibility. The differences in heights between the two hands and the comparisons of the sensations in the arm and attached muscles would allow users to feel the relationships between the curves.

\subsection{Parallel processing}

The authors envision a program that would integrate with cellphones and send messages when simulations have finished. From here, researchers could update arguments and rerun models through voice-recognition similar to sending a text message. We anticipate that some aspects of modeling can be done hands-free and remotely from a cell phone.

\subsection{Teaching}

Histograms are a standard data visualization tool. Teaching this plot involves discussing an appropriate bin width. With current technology learners plot multiple plots to discern an appropriate range for this argument. A better teaching tool might be developed using a haptic glove where the distance between the thumb and forefinger align with the bin width. The bin width of the histogram would increase simultaneously with the distance between the fingers. This mimics the growth of the bin and allows for somatic understanding of what the argument does that is impossible to achieve by solely changing the argument in a function. This tool would engage students who struggle to understand this and deepen the understanding of others.
 
An extension of the histogram is the density plot. Like the histogram, drawing this plot involves the bin width argument for which the authors suggest a glove. In addition to the bin width argument there is a kernel argument that could be voice activated. If the output is a smooth curve---say, from a Gaussian kernel---sonifying the resultant density function could deepen the understanding of the learners and engage learners who might not understand. When the bin width is too small the pitch of the sonification would vary widely; when it is too big it would be overly simple. 

In both of the above examples 3D printing would also deepen understanding. For the histogram, a 3D dot plot of the values that can be layered onto multiple 3D histograms with different bin widths would add an extra dimension. Density curves could also be 3D plotted.  

An even more fundamental example is the mean. The mean is also the center of mass for a distribution which could be built with 3D printing. Learners would have a deeper understanding of the importance of this computation.

Correlations could be explored using a virtual reality (VR) headset. A scatterplot where the axes can be shifted from the origin to the means of $x$ and $y$, then the squares for the numerator drawn with different colors for negative and positive values. Seeing multiple data sets used to iterate the numerator would allow learners to deeply understand the reasoning behind the formula in a way that is difficult.

Iterative algorithms like the expectation-maximization algorithm are similar to walking up a lighthouse. We could demonstrate this with a VR headset. The observed data log-likelihood after convergence would be visible at the top. The expectation step can be seen on one side of the lighthouse and maximization on the opposite. For people with low vision, the distribution could be 3D printed. This reinforces the dependence of the current distribution on prior iterations in a way that is difficult without these somatic cues. This could be extended to explore different initial values by changing the location of the initial step. Additionally, the number of iterations can vary between viewings, or controlled through voice-recognition software.

The authors take a broad view of learners. Data scientists often need to explain concepts to stake holders and we anticipate that these tools could be used outside of a traditional classroom.

\subsection{Using pitch, volume, and timbre to convey statistical information}

The past two years have been altered by COVID-19. Several months into the pandemic there were a number to tests available to people but understanding sensitivity, the true positive rate, and specificity, the true negative rate, could be communicated multiple ways besides stating the definition. Sonification could be explored. An octave is familiar to most ears. To sonify testing data we could start with a note and use the octave increase for 100\% sensitivity, and a decrease of an octave for 100\% specificity. The actual sensitivity and specificity notes would be prorated to the actual value; we wouldn't expect the actual octave since tests with 100\% sensitivity or specificity are uncommon. In addition, we could use a glissando (glide) sound starting at the original then rising to the sensitivity level, return to the original tone, then glissando down to the specificity. As different tests take different amounts of time to process, the speed of the glissando could be related to the processing time, with quicker tests having shorter ``songs'' and vice versa. To extend this, multiple tests could be considered with dependence in the calculations of sensitivity and specificity and fixed speed of the glissando.

Another opportunity to sonify COVID relates to the differences in rates of change in different countries. \cite{soundRates} explores pandemic graphs, and one plot, ``COVID-19: How many days it took to reach 500 deaths?'', could be sonified using glissandos. Each group of 500 deaths could be a one-octave glissando with duration related to the time taken for a country to reach that iteration. Timbre should be chosen carefully to reflect the seriousness of the topic.

As the size of data increases, variable selection has increased in importance. Lasso and ridge regression are both methods that can be used to fit multiple models each with different penalties (lasso is the sum of the squares of the coefficients, ridge is the sum of the absolute values of the coefficients). There are a number of ways to vivify variable selection. The resulting plot from ridge or lasso regression of coefficient estimates vs.\ the regularization parameter $\lambda$ could be sonified, 3D printed, or seen through a VR headset. For sonification, each variable could be mapped to a different timbre dependent on the type of variable. For one-variable data, astronomers visualize x-rays (high-energy photons) with blueish purple which is the highest frequency of visible color. To map to timbre, the xylophone is the instrument with the highest timbre so it is mapped to the highest-energy photons. Following this trend, violins is used for near-infrared light, and soft piano for infrared light \citep{Arcand}. A similar process could be used for multiple variables; synesthesia might be helpful.
 
As $\lambda$ becomes large, all coefficients go to zero due to the penalty. For this reason, the authors suggest reading the plot from right to left instead of the normal left to right. All variables/timbres/instruments start on the same tone which represents 0 and a high value of $\lambda$, as time passes, the pitches adjust with the values of the coefficients as the value of $\lambda$ decreases. The sonification ends with the coefficients from the least squares fit sonified when $\lambda$ is 0. Users should have the ability to choose the variables they would like sonified to simplify the sound if they desire.

In terms of a VR headset, the authors again recommend reading the plot from right to left and enlarging the plot a considerable amount. Visualizing each coefficient path as a tunnel wherein the user can travel from a large $\lambda$ to a $\lambda$ of 0 could increase understanding. One day it might be possible to experience it like an amusement park ride. 

\section{Discussion}
\label{sec:conc}

\subsection{First steps}
One can start by applying existing technology to tasks of statistical data analysis,  For example, it should be possible to build a digital signal processing \cite{ThinkDSP} package in R to enable  works like those done at System Sounds to be automated in R  \citep{sonification, systemSounds}. Sonification is a natural extension of data visualization for low-vision and blind people.  The next step would be to develop data vivifications that include both sight and sound. Some preliminary work has been completed in this field \citep{StormsRussellL2000IiPQ,2011Tsh} but more is needed.

Simultaneous to this, it would make sense to develop a network of people within the low-vision and blind community to support this process. To achieve the goal of making data more accessible quickly, listening to those who have been continually disenfranchised is imperative. Technical developments should be done in concert with this group to ensure that the tools being built, and the order in which they are being built, are optimized to meet the needs of the community. This would build upon community-based research initiatives where participants guide researchers towards a shared goal.

\subsection{Using theoretical ideas to evaluate and understand particular data vivifications}

Chernoff reduced the multiple complexities of faces to a few characteristics like length of the nose and curvature of the mouth. To extend his method, \cite{Chernoff} suggested ``adding ears, hair, [and] facial lines.'' The abstract of the paper uses the term ``cartoon,'' and humans have evolved to read people not cartoons. Chernoff did not understand the perceptual properties of faces before trying to create them \citep{CF_bad}. This method simplified the human Umwelt of a face.

It is clear to us that Arabic numerals are much easier to manipulate than Khipu. Khipu were not calculators so extensions of this method to basic arithmetic was not possible \citep{khipu_bad}. This method was too complex to be extended.

To summarize, Chernoff faces simplify the human Umwelt without understanding them, and Khipu is not simple enough to extend. Twenty years from now, data vivification methods will likely include methods that are simple enough to be extended, as well as complicated enough to represent the human Umwelt. This is not to say that Chernoff's work is valueless; we often need to try out ideas and see what does not work in order to better understand how to do better. 

An exception to this is the famous image by Charles Minard of Napoleon's armies diminishing as they marched through Russia. This mode of display is not extendable or applicable in many other settings---it takes advantage of a particular data structure that allows multiplexing of time and two spatial dimensions---but many experts hold it in high regard as a model of data storytelling.

Lack of diversity in technical fields has had negative consequences on what data are collected and on the development of algorithms \citep{Compas, pmlr-v81-buolamwini18a, AutoIneq,  racialBiasOxygen}. This goes beyond racism and sexism; the community also lacks members with physical constraints such as restricted sight and hearing. This impacts all aspects of the data pipeline, including perhaps a problem of integrating communication and visualization with other aspects of statistics. Through delivering data differently, the authors hope to foster a more inclusive environment throughout science.

\subsection{Engaging the open-source software community}

A large community exists around the open-source statistical programming language R. The preeminent conference, useR! 2021, has multiple tutorials available in languages other than English. In addition, the R Consortium sponsors over 92 R User Groups in over 38 countries with over 60,000 members. R Forwards is another community whose mission is ``leading the R community forwards in widening the participation of women and other under-represented groups.'' R Ladies is  group that works towards diversity in the R community; at publication, there are over 75,000 members in 57 countries. The community is focused on including those from diverse backgrounds and experiences. One focus of The Minority in R community is accessibility. Several members are blind and others have extensive backgrounds in accessibility and are presenting and teaching the topic.

Although the RStudio interactive development environment is not yet accessible, using R through emacs is accessible for those with low vision  \citep{JSSv058s01}. Our second author confirms that, despite being relatively unfamiliar with R, he readily embraced it after publishing a comparative analysis of R, SAS, SPSS, and Minitab for people with low vision \citep{JSSv058s01}.  

It is important to engage this community about delivering data differently as part of a long conversation that will not be limited by today's thinking. 

\subsection{Replacements and improvements}
We tell our children, ``Don't try to be someone else.  Just be the best `you' you can be.''  The same goes for the methods discussed in the present article:
\begin{itemize}
\item Sonifications in the form of music can be emotionally engaging and can also be processed in the background, unlike visualizations, which require conscious attention to convey their information.
\item The haptic sense is inherently interactive, as it is through pushing that you feel the resistance in your muscles.
\item Through the sense of touch you can perceive three dimensions directly with no need for projections.
\item An effective voice dialogue has no need to pass a Turing test; it is more effective when it is transparent enough that the user can work directly with its algorithmic nature.
\end{itemize}
The data-exploration tools we would like to develop can serve as replacements for visual tools for blind users, but they also have their own unique advantages.  Thus, when considering vivification of data or statistical models using sound or touch or the muscular resistance sense, let's try to go directly from the statistical or scientific problem to the sense in question, rather than first ``translating'' to a visual display and then ``translating'' to the non-visual mode.

\subsection{Looking forward}
Technology can expand the human Umwelt \citep{SensRev}. The scientific term for this is sensory augmentation. Skin is a particularly good at sensing different patterns and pressures, and there has been research on building sound-to-touch interfaces that work without lipreading \citep{NovichScottD2015Usat}. The future is likely to include interactions with computers far beyond our imagination as we learn to develop new skills. They will likely not be limited to keyboard and mouse; indeed, those technologies may go the way of the rotary phone and typewriter. 

Computer scientists will benefit from the support of consultants with advanced somatic skills to imagine new methods of interaction. The seminal work The Body Keeps the Score discusses the consequences of trauma upon the brain \citep{TBKS}. Few people make it through life without times of loss of safety; imagine the results of  such intrusions on your body. Scientists believe that if the body feels safe the brain heals \citep{Heal}, and prioritizing feelings of safety could be one way forward. Methods that assess how technology impacts upon the human body will support this work.

In addition, consider the expansion of of nascent technologies, and technologies not yet developed, to include statistical methodologies. Currently, work on visualization and vivifications focuses on data display, but this is only part of statistical or data science workflow. A first step going forward would be to identify a set of example problems in applied statistics and explore implementations of different sorts of vivifications. Once some examples have been made available, it should be possible to think of how to integrate a range of vivification tools into data science workflow. Given that it took many years to incorporate visualization into applied statistics, we do not anticipate this will be easy, but we think the way forward is to take advantage from all the available technology (some of which has been reviewed in the present article) and think about how to apply them to various challenges of real-world data analysis. 

We believe that future delivery of data will stimulate more of the human Umwelt then current delivery methods. Research suggests that high quality audio increases perception when coupled with high quality visualizations  \citep{StormsRussellL2000IiPQ,2011Tsh}. This has the potential to increase accessibility for those with low vision and increase perceptiveness for all.  We would like to see this happen faster and would like those with statistical training to take leadership roles.

Scientific misunderstandings abound even among the well educated \citep{PrivUni}.  Indeed, misconceptions are an important part of the learning process as students integrate new knowledge \citep{Sadler98}. One way to address this knowledge is data vivification where students have the opportunity to integrate new knowledge using multiple aspects of their Umwelt \citep{SCHNEPS2014}.

Vaccines only work if communities take them. With the rise of mistrust in science, scientists must carefully consider the limitations of the impact of their work if communities do not believe them. Research suggests that delivering data differently may mitigate some of these issues. We encourage scientists to learn about and utilize these approaches in reaching out to their communities.

In the words of \cite{translationalSkills}, ``the most important skill for a translational scientist [is] to be able to work in a collaborative environment with people from other fields. As a biostatistician, having this skill is not a choice. You have to work with doctors and scientists to solve problems. You must communicate across disciplinary boundaries, earn trust and learn to speak your colleagues' specialized language. Over time, you discover what is important to your collaborators and understand why they work a certain way. For example, the cardiologists I [Roland Matsouaka] work with not only have training in medicine but are informed by constant interaction with patients. Their perspective reminds me that my job is not just to write complicated equations---I also need to translate those equations into a common language that people can understand.''  We see our work as extending the work of translational science within scientific practice and among the general public.

\subsection{Win conditions}

We see two success criteria. Most directly, we would like for blind and low vision people can more easily access, experience, and work with.  Beyond that, we would like to develop vivifications that make statistical workflow more effective for sighted people by making use of the unique benefits that other senses have to offer.  This should not be controversial; consider, for example, the familiar observation that learning and memory are enhanced when multiple senses are involved, as when we can remember a particular formula by visualizing where it was on the blackboard when it was presented in a lecture, or when a long train of thought is triggered by an unexpected smell or a sudden change in temperature.  \cite{stilgoe1998} discusses the way that even familiar landscapes convey unexpected information and stimulate creative thought, if we train ourselves to focus on details.  Vivifications should be a way to do this. The authors would like to develop tools that will continue to be useful and be developed, not to solely go viral and then be forgotten. The future of our current practice of machine learning, data science, and artificial intelligence is what we would like to develop and explore and what we are addressing in this article, and one challenge going forward is to consider how to measure the success of this endeavor.

\bibliography{citations}

\end{document}